# Validation of the multi-mission altimeter wave height data for the Baltic Sea region

Nadezhda A. Kudryavtseva and Tarmo Soomere

Institute of Cybernetics at Tallinn University of Technology, Akadeemia tee 21, 12618 Tallinn, Estonia; nadia@ioc.ee

**Abstract.** We present a complete cross-validation of significant wave heights (SWH) extracted from altimetry data from all ten existing satellites with available in situ (buoy and echosounder) wave measurements for the Baltic Sea basin. The main purpose is to select an adequate altimetry data subset for a subsequent evaluation of the wave climate. The satellite measurements with the backscatter coefficients >13.5 cdb, errors in the SWH normalized standard deviation >0.5 m and snapshots with centroids closer than 0.2° to the land are not reliable. The ice flag usually denotes the ice concentration of >50%. The presence of ice affects the SWH data starting from concentrations 10%, but substantial effects are only evident for concentrations >30%. The altimetry data selected based on these criteria have very good correspondence with in situ data, except for GEOSAT Phase 1 data (1985–1989) that could not be validated. The root-mean-square difference of altimetry and in situ data is in the range of 0.23–0.37, which is significant for the Baltic Sea, compared with an average wave height of ~1 m. The bias for CRYOSAT-2, ERS-2, JASON-1/2 and SARAL data is below 0.06 m. The ENVISAT, ERS-1, GEOSAT and TOPEX satellites revealed larger biases up to 0.23 m. The SWH time series from several satellite pairs (ENVISAT/JASON-1, SARAL/JASON-2, ERS-1/TOPEX) exhibit substantial mutual temporal drift and part of them evidently are not homogeneous in time. A new high-resolution SWH data set from the SARAL satellite reveals a very good correspondence with the in situ data and with the data stream from previous satellites.

## INTRODUCTION

The complexity of wave fields and particularly their spatial and temporal variations in the Baltic Sea extend far beyond the typical features of surface wave climate in water bodies of comparable size (Leppäranta & Myrberg 2009; Soomere 2016). Owing to the relatively small size of the sea, the wave fields mainly follow changes in the wind properties and at times reveal even regime shifts (Soomere & Räämet 2014). The processes that depend on wave activity are thus particularly susceptible to wind climate changes (Tõnisson et al. 2011; Suursaar et al. 2014). The elongated shape of the sea and some of its subbasins augments the effects related to the rotation of strong wind directions (Viška & Soomere 2012; Pindsoo & Soomere 2015). The seasonal presence of ice adds complexity to the system. As ice cover protects the nearshore domains and coasts during a large part of the windy season, its reduction via the climate warming may lead to a major increase in the wave energy approaching the nearshore. Changes in ice cover may, thus, substantially modify the evolution of the coastal environment (Orviku et al. 2003; Ryabchuk et al. 2011).

The impact of the listed features is interspersed with other elements that enhance the intrinsic complexity of the Baltic Sea wave fields. The presence of extensive archipelago areas (Tuomi et al. 2014), relatively shallow regions and occasional convergent wind patterns (Soomere et al. 2008) gives rise to significant spatio-temporal variability in the wave properties (Soomere & Räämet 2011). Specific features of wave generation near irregular coastline sections (Kahma & Calkoen 1992; Tuomi et al. 2012) or the impact of so-called slanting fetch conditions (Pettersson et al. 2010) make the quantification of wave fields even more difficult. The evolution of the southern and eastern partially subsiding sedimentary coasts is almost fully wave-dominated (Harff & Meyer 2011; Zhang et al. 2013; Deng et al. 2014) and any considerable change in the wave properties may substantially impact their development and even stability. In this context, proper quantification of the Baltic Sea wave climate is a hugely important and challenging task.

Currently, three sources of wave data are available for the research community, namely, direct measurements, modelling and satellite altimetry. The direct measurements (usually waverider buoys but also echosounder-or ADCP-based devices) have provided high-quality, detailed variability of the major features of wave fields such as significant wave heights (SWH), wave periods and directions since the mid-1970s. These data are regularly reviewed (e.g., Pettersson et al. 2013) and are one of the core assets in the evaluation of the local wave climate in different domains of the Baltic Sea (Broman et al. 2006; Tuomi et al. 2011; Soomere et al. 2012; Suursaar 2013, 2015).

However, the cost of direct measurements does not allow covering the whole area of interest. Wave measurements in the northern regions of the sea are limited to the ice-free time (Tuomi et al. 2011). As a consequence, the relevant data sets have long gaps and several commonly used characteristics (e.g., annual mean wave height or average period) may become meaningless (Tuomi et al. 2011; Ruest et al. 2016). Moreover, extensive spatial variations in the Baltic Sea wave climate that are often almost uncorrelated (Soomere & Räämet 2011, 2014) suggest that direct measurements result in very limited knowledge.

Wave modelling is performed on large spatial scales and provides good results for open ocean conditions. The limitation of modelling is that the results are dependent on wind quality and (to a lesser extent) assumptions fed into the model. Even though the increased quality of wind fields has significantly improved the accuracy of wave hindcast and forecast in the Baltic Sea (Cieślikiewicz et al. 2008; Tuomi et al. 2011, 2014), detailed reconstruction of wave climate is still a major challenge for this basin (Hünicke et al. 2015). Along with the problems associated with the variability of ice cover extension from 12.5% to 100% of the Baltic Sea (Leppäranta & Myrberg, 2009), modelling efforts still suffer from major mismatches between different model outputs and between modelled and measured wave data. Decadal and multi-decadal reconstructions of wave fields (Soomere & Räämet 2011; Tuomi et al. 2011; Alari 2013) show qualitatively similar patterns but reveal significant mismatches in quantitative terms (Nikolkina et al. 2014). The most probable reason for the mismatches is that the wind data sets tend to lose their quality at some distance from the shore (Räämet et al. 2009).

In such a situation remote sensing data, first of all, systematic exploitation of properties of the wave field derived from satellite altimetry, may provide a sensible solution. Various remote sensing methods cover large spatial areas and provide fairly (albeit not completely) homogeneous and continuous (along a certain line) data about the sea state on the large scale. Multiple studies have focused on cross-validation of the satellite altimetry data with the buoy data. The high quality of the altimeter data has been demonstrated both globally (Queffeulou 2004; Queffeulou et al. 2010; Queffeulou & Croizé-Fillon

2012) and for smaller basins, e.g., for the Chukchi Sea (Francis et al. 2011), Indian Ocean (Shaeb et al. 2015; Kumar et al. 2015), Mediterranean Sea (Cavaleri & Sclavo 2006; Galanis et al. 2012) and Arabian Sea (Hithin et al. 2015).

The observations in question span today over more than 20 years and thus provide a valuable contribution not only to evaluate the sea state at present but also to studies of wave climate. The use of satellite altimetry has substantially contributed to the relevant analysis in remote locations of the open ocean. For example, Young et al. (2011) found a significant global increase in wind speed and wave height. The trends are, however, different in different parts of the ocean and for various parameters of wave fields. The North Atlantic region exhibits a small decrease in the mean SWH, but a definite increase in its 90th and 99th percentiles, which displays an increase in the severity of the storms. A significant rise in the wave height by 20 mm/yr was identified in the Chukchi Sea (Francis et al. 2011), probably because of the shrinking of the ice cover in the region. The annual mean SWH in the Central Arabian Sea exhibits a positive trend of 6.3 mm/yr (Hithin et al. 2015).

The use of satellite altimetry is problematic in coastal areas and partially ice-covered water bodies, and none of the mentioned studies included the Baltic Sea. The use of this approach has been scarce in this region. The relevant studies include only comparisons of model results with TOPEX/POSEIDON altimetry data in 1996–1999 (Cieślikiewicz et al. 2008) and with the JASON-1 altimetry for five years (Tuomi et al. 2011). Recently, Rikka et al. (2014) used high-resolution SAR data for the evaluation of wave properties at a short distance (a few hundreds of metres) from the shoreline.

Large discrepancies between the SWH values provided by different missions and time drifts of single missions (Queffeulou 2004; Queffeulou & Croizé-Fillon 2012) are generic problems with the use of satellite altimetry. This concerns first the early missions. For example, a bias of about 0.5 m is observed between ERS-1 and TOPEX in terms of global ocean monthly mean (Queffeulou et al. 2010). An effort to provide corrected SWH data sets has been performed within the European Space Agency Globwave project (www.globwave.org). These biases and shortages contaminate the particular and average values of SWH in single locations and may introduce fake trends of changes in the basin-scale average SWH. However, they apparently do not affect the qualitative appearance of spatial patterns of wave intensity as well as changes in these patterns.

Another possible reason for such an infrequent use of altimetry data in the Baltic Sea is the general belief that this method of retrieving wave properties has too large uncertainties in such small and seasonally ice-covered sea areas where almost the entire sea surface can be considered as a nearshore region. In this paper, we demonstrate that these limitations for the use of satellite altimetry in the Baltic Sea basin can be circumvented by careful analysis of the geometry of the basin, ice conditions and spatial coverage of each altimetry snapshot. As a side result, a thorough validation of the entire data set of 30 years (1985–2015) of multi-mission altimetry covering the whole of the Baltic Sea makes it possible to identify the magnitude of mismatches between in situ and satellite data and between different missions, and possible temporal drifts in data provided by these missions for the Baltic Sea basin. More generally, this analysis allows us to identify these data sets that provide adequate information. Some preliminary results of the study have been presented in Kudryavtseva et al. (2016). To assess the limitations of the satellite altimeter data quality, the data were cross-matched with available direct wave measurements, mainly with the information from buoys of the Swedish Meteorological and Hydrological Institute and Finnish Meteorological Institute.

## DATA AND METHODS

### Altimetry data

The satellite altimetry data used in this paper were obtained from the Radar Altimeter Database System (RADS) database (http://rads.tudelft.nl/rads/rads.shtml) (Scharroo 2012; Scharroo et al. 2013). This database provides altimeter data uniformly reduced for multiple missions, which reduces a bias between data from different satellites and makes it possible to examine long-term changes in the wave climate. As mentioned above, the RADS data still reflect deviations between different missions and time drifts of single missions (Queffeulou 2004; Queffeulou & Croizé-Fillon 2012). For the Baltic Sea region, the relevant data from nine satellites are available over the period of 1985–2015 and analysed in this paper (Table 1). The satellite altimeters measure significant wave heights but provide no information about the wave period and direction. The basic data of these missions can be retrieved from, e.g., *JASON-2 Handbook* (JASON 2008), Vignudelli et al. (2011), *CRYOSAT Product Handbook* (ESA 2012) or *SARAL/AltiKa Products Handbook* (Bronner et al. 2013).

Even though different satellites have a different density of observations, the measurements fairly densely cover the entire Baltic Sea (Fig. 1). The SARAL and CRYOSAT-2 altimeters have the highest density of observations over the Baltic Sea region, with roughly daily visits to this basin. On the contrary, the POSEIDON altimetry data have very sparse coverage and often contain only 1–2 measurements per month.

**Table 1.** Satellite missions that provide altimetry data for the Baltic Sea region. See the descriptions of bands in Vignudelli et al. (2011)

| Satellite | Time of observations | Band |
|---|---|---|
| CRYOSAT-2 | 2010–2015 | Ku |
| ENVISAT | 2002–2012 | Ku,S |
| ERS-1 | 1991–1996 | Ku |
| ERS-2 | 1995–2004 | Ku |
| GEOSAT | 1985–1989 | Ku |
| | 2000–2008 | Ku |
| JASON-1 | 2002–2013 | Ku,C |
| JASON-2 | 2008–2015 | Ku,C |
| SARAL | 2013–2015 | Ka |
| TOPEX | 1992–2005 | Ku,C |
| POSEIDON | 1992–2002 | Ku |

We focus on the analysis of Ku-band altimeter data for all satellites except for SARAL, for which only Ka-band is available. The data were selected in the range of latitudes 53.5 to 67°N and longitudes 6 to 30°E over the Baltic Sea (land/sea flag4 = '0').

### Data from buoys

For the validation of the altimeter data, we use SWH data from waverider buoys (Fig. 2, Table 2) of the Swedish Meteorological and Hydrological Institute (SMHI) and the Finnish Meteorological Institute (FMI) and older data from upward-looking echosounders at Almagrundet (Martensson & Bergdahl 1987; Broman et al. 2006). Data from the Swedish Meteorological Institute were retrieved through the open data interface (http://opendata-download-ocobs.smhi.se/explore/). The FMI open data (en.ilmatieteenlaitos.fi/open-data-manual) were accessed through the data mining tool described in Giudici & Soomere (2015). The buoys used for the comparison were selected based on data availability and a duration of observations longer than a year (to ensure that the yearly averages and the following statistical analysis make sense). Some of the FMI buoys cover longer time intervals of observations (Pettersson et al. 2010, 2013; Tuomi et al. 2011), but we limit our analysis to the available open data.

According to the manufacturers, contemporary directional waveriders (including those used by the SMHI and FMI) retrieve the wave height for quite a wide range of wave periods (1–30 s) with an accuracy of 0.5% of measured values. This accuracy is usual in field conditions where the typical correlation coefficient of readings of closely located measurement devices is 0.97–0.99 (e.g., Taylor et al. 2002; Liu et al. 2014) and where small variations in the SWH at different locations are an intrinsic feature of natural sea states. The presence of moorings may decrease this accuracy to some extent (Liu et al. 2014). The estimates of SWH at Almagrundet are derived using the 10th largest wave during the measurement interval of 640 s and the assumption that wave heights are Rayleigh distributed (Broman et al. 2006). This assumption considerably increases the uncertainty of the readings, and the deviation between such estimates of SWH and alternative estimates from the wave spectrum reaches 7% for extreme sea states (Broman et al. 2006).

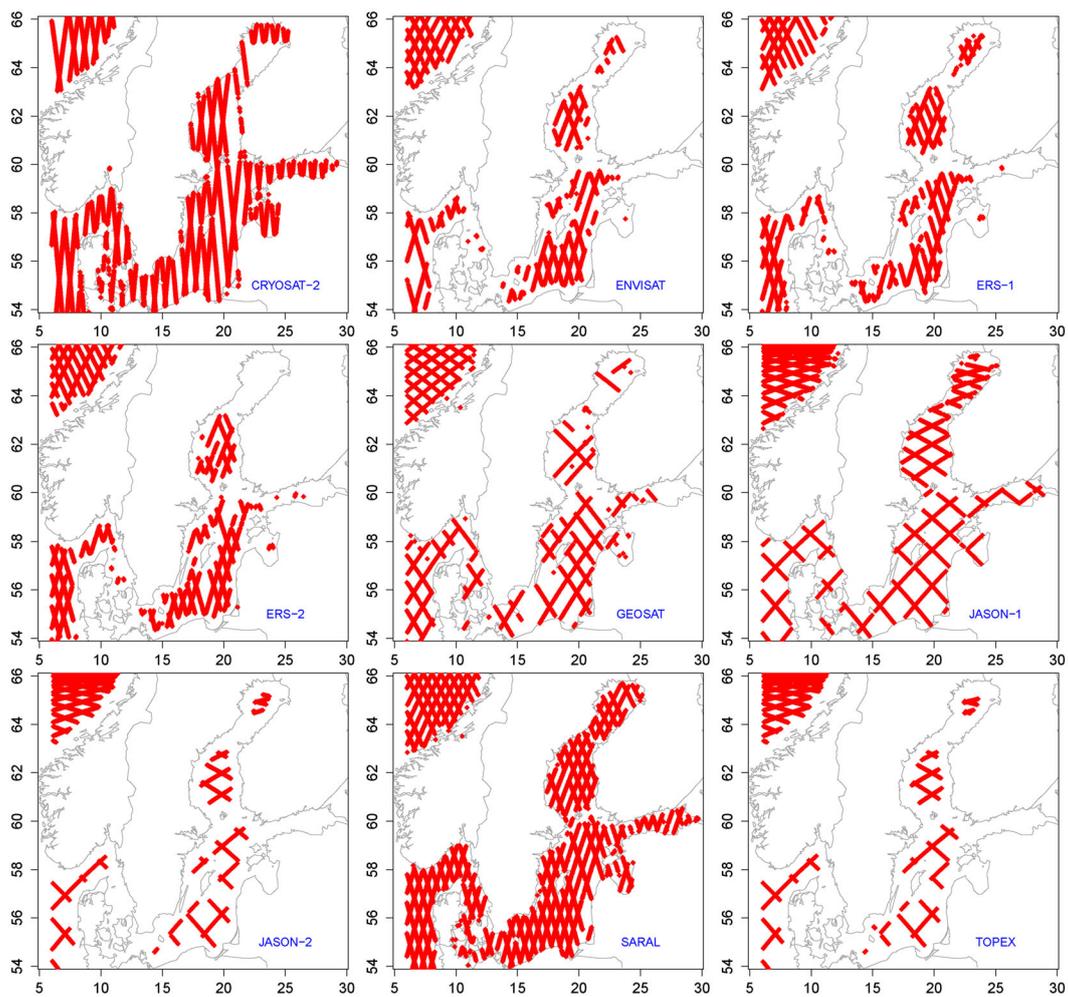

**Fig. 1.** Altimeter tracks during one month over the Baltic Sea. Latitude is shown at the *y*-axis and longitude at the *x*-axis.

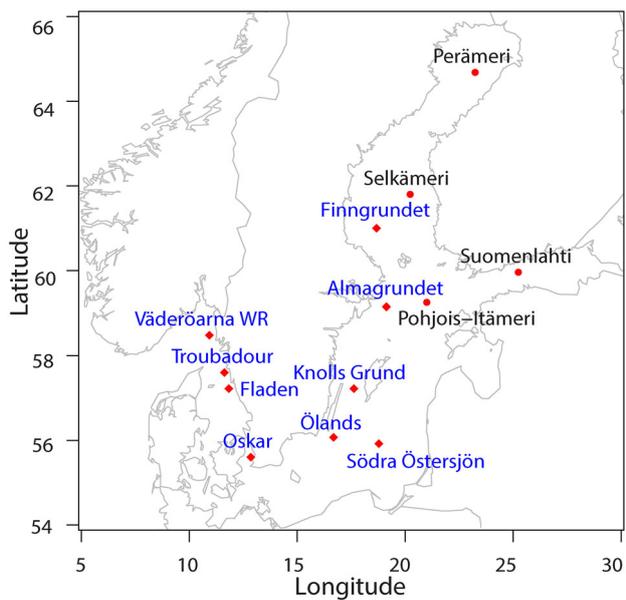

**Fig. 2.** Location scheme of the Baltic Sea and wave measurement locations used in the analysis. The SMHI stations are marked with diamonds, the FMI stations with circles.

**Table 2.** The geographical coordinates, time intervals and distance from the land $D$ of observations at the measurement sites, the data of which are used in this paper. Note that at Almagrundet the measurements were performed using an upward-looking echosounder attached to the seabed in about 30 m deep area (Broman et al. 2006)

| Buoy | Lat, °N | Lon, °E | Time of observations | $D°$ |
|------|---------|---------|----------------------|------|
| Almagrundet (SMHI) | 59.15 | 19.13 | 1978–2003 | 0.20 |
| Finngrundet (SMHI) | 61.00 | 18.67 | 2006–2015 | 0.54 |
| Fladen (SMHI) | 57.22 | 11.83 | 1988–1999 | 0.17 |
| Knolls Grund (SMHI) | 57.22 | 17.62 | 2011–2015 | 0.37 |
| Ölands Södra Grund (SMHI) | 56.07 | 16.68 | 1978–2004 | 0.25 |
| Oskar (SMHI) | 55.60 | 12.85 | 1983–1999 | 0.04 |
| Perämeri (FMI) | 64.68 | 23.24 | 2012–2014 | 0.62 |
| Pohjois-Itämeri (FMI) | 59.25 | 21.00 | 1996–2015 | 0.63 |
| Selkämeri (FMI) | 61.80 | 20.23 | 2011–2015 | 1.10 |
| Södra Östersjön (SMHI) | 55.92 | 18.78 | 2005–2011 | 1.12 |
| Suomenlahti (FMI) | 59.96 | 25.24 | 2000–2015 | 0.22 |
| Troubadour (SMHI) | 57.60 | 11.63 | 1978–2003 | 0.09 |
| Väderöarna WR (SMHI) | 58.48 | 10.93 | 2005–2015 | 0.31 |

**Cross-matches between satellite and buoy measurements**

To validate the SWH data provided by satellite altimeters, we compare them to the available buoys data. Following Høyer & Nielsen (2006), we look for satellite altimeter data centred within 0.3° distance from the buoy and time difference between the instants of observations less than 30 min for each buoy measurement. The cross-matching radius in degrees is the same along latitudes and longitudes. This means that the cross-matching area is a rectangle with a North–South extension (about 100 km) about twice as large as the East–West extension (about 50 km). The analysis of SWH residuals does not show any dependence on the time difference and spatial difference for this set of values. This is consistent with the results of Monaldo (1988) and Zieger et al. (2009) who showed that half an hour time separation leads to the uncertainty of 0.3 m for the SWH.

As the TOPEX satellite does not have any data to 0.3° from the buoys, the cross-matching radius for this mission was increased to 0.4°. The POSEIDON satellite has no data within 0.5° from the buoys. As the use of even longer cross-radii is not justified in spatially fairly inhomogeneous wave fields of the Baltic Sea, the POSEIDON data were excluded from the following analysis.

The resulting samples were further scanned for possible errors or other shortages. We first excluded the data with the backscatter coefficient >13.5 cdb (see, e.g., Gairola et al. (2014) for a description of this quantity) and large errors in SWH normalized standard deviation. These levels of the backscatter coefficient correspond to low wind speeds of <2.5 m/s. As the typical significant wave heights of the Baltic Sea are around 1 m (Tuomi et al. 2011; Soomere 2016), potential errors larger than half of this value (>0.5 m) are generally unacceptable.

Several phenomena may substantially affect the outcome of satellite altimetry. To avoid the impact of such phenomena, we used flags provided by each mission, which marked data as bad because of rain or the presence of sea ice. We also applied similar flags that warned about possible large errors in the range, backscatter coefficient and significant wave height (flags 7, 11, 12, 13 in the RADS database). On top of that, it is well known that the GEOSAT data have a problem with the attitude status (e.g. Sandwell & McAdoo 1988). For this reason, the attitude flag (flag 1 in the RADS database) was applied to these data as well. JASON-1 and ERS-1 missions have a few measurements with SWH equalling exactly zero. As such measurements are probably simply erroneous, these data were removed as well.

The ERS-2 and TOPEX satellites showed significant scatter in comparison with the Almagrundet echosounder measurements. Further analysis indicates that the enhanced level of scattering was mostly owing to large differences between the altimetry and echosounder data within a short time range December 1996–April 1997. These months correspond to a data subset that was recorded by the WHM device just before a longer gap in the Almagrundet data in 1998. As the quality of the WHM data from Almagrundet is problematic and the recorded wave periods are unrealistic (Broman et al. 2006), the mismatch and the resulting scatter most likely are caused by problems with the echosounder data. Interestingly, if this period is excluded, the rest of the data do not exhibit any significant bias between the ERS-2 and TOPEX observations and the echosounder recordings. This feature suggests that the overall cyclic variation in wave heights at Almagrundet (Broman et al. 2006), and particularly a decrease in the wave heights in 1999–2003, is realistic.

The analysis of the cross-matches reveals a good correspondence between the buoy recordings and wave heights extracted using satellite altimetry (Figs 3, 4). The quality of cross-matching demonstrated for JASON-1 is at the same level for other satellites. The best match is at locations in the offshore of the Baltic Proper, somewhat less convincing at measurement sites in smaller sub-basins such as the Bay of Bothnia or the Gulf of Finland and poorer in nearshore locations such as Troubadour.

To examine if there is a bias between altimeter data and buoys, we fit a reduced major axis regression in cross-matches between a single satellite and all buoys. The relevant parameters are calculated with 'lmodel2' package in R statistical computing language (R Core Team 2015). The fitted intercept for JASON-1 SWH versus cross-matches with all buoys (incl. the echosounder at Almagrundet) is $-0.001 \pm 0.02$ (Fig. 5). The level of the relevant $p$-value of 0.5 indicates that the nonzero location of this value is not statistically significant.

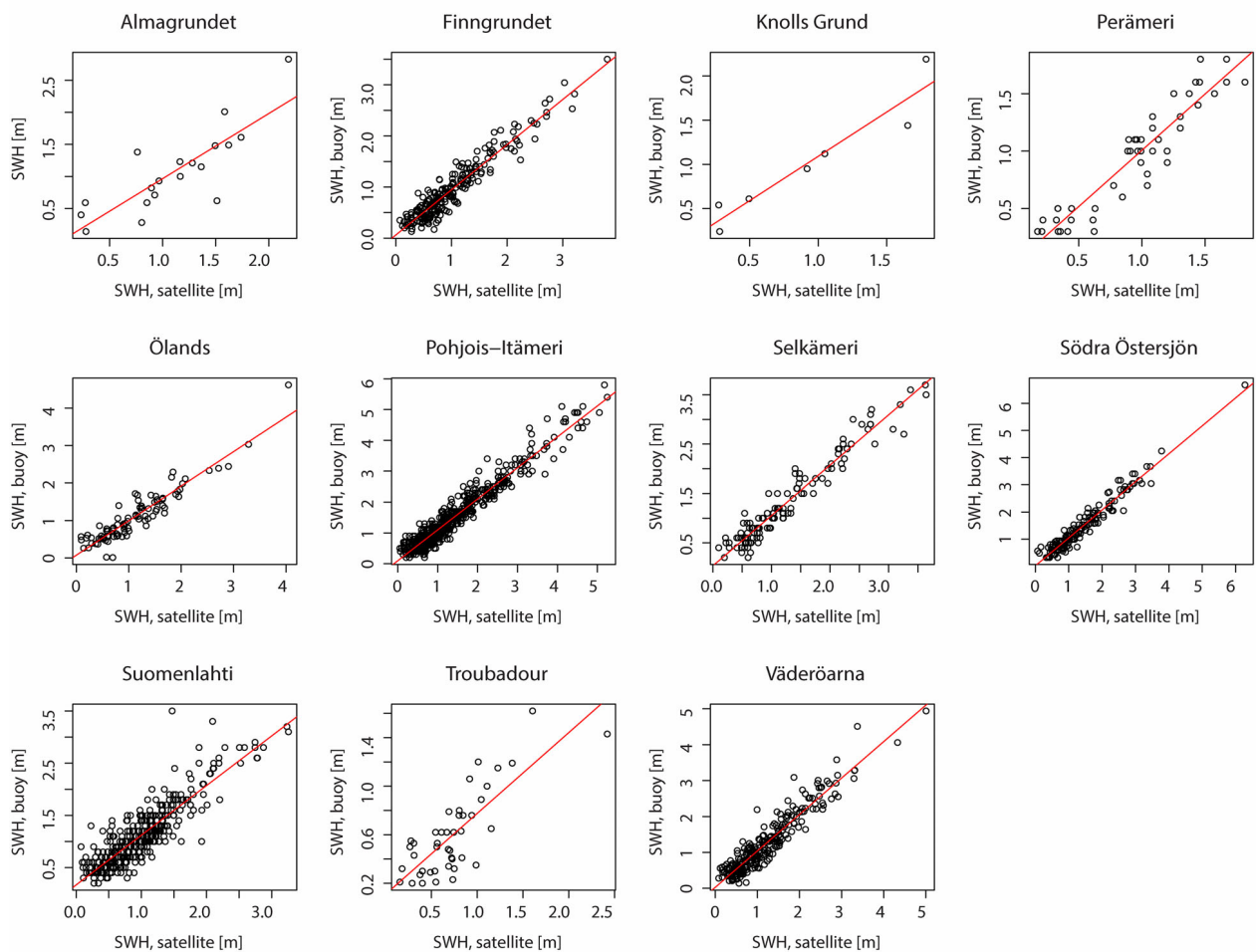

**Fig. 3.** JASON-1 significant wave heights versus in situ measured significant wave heights.

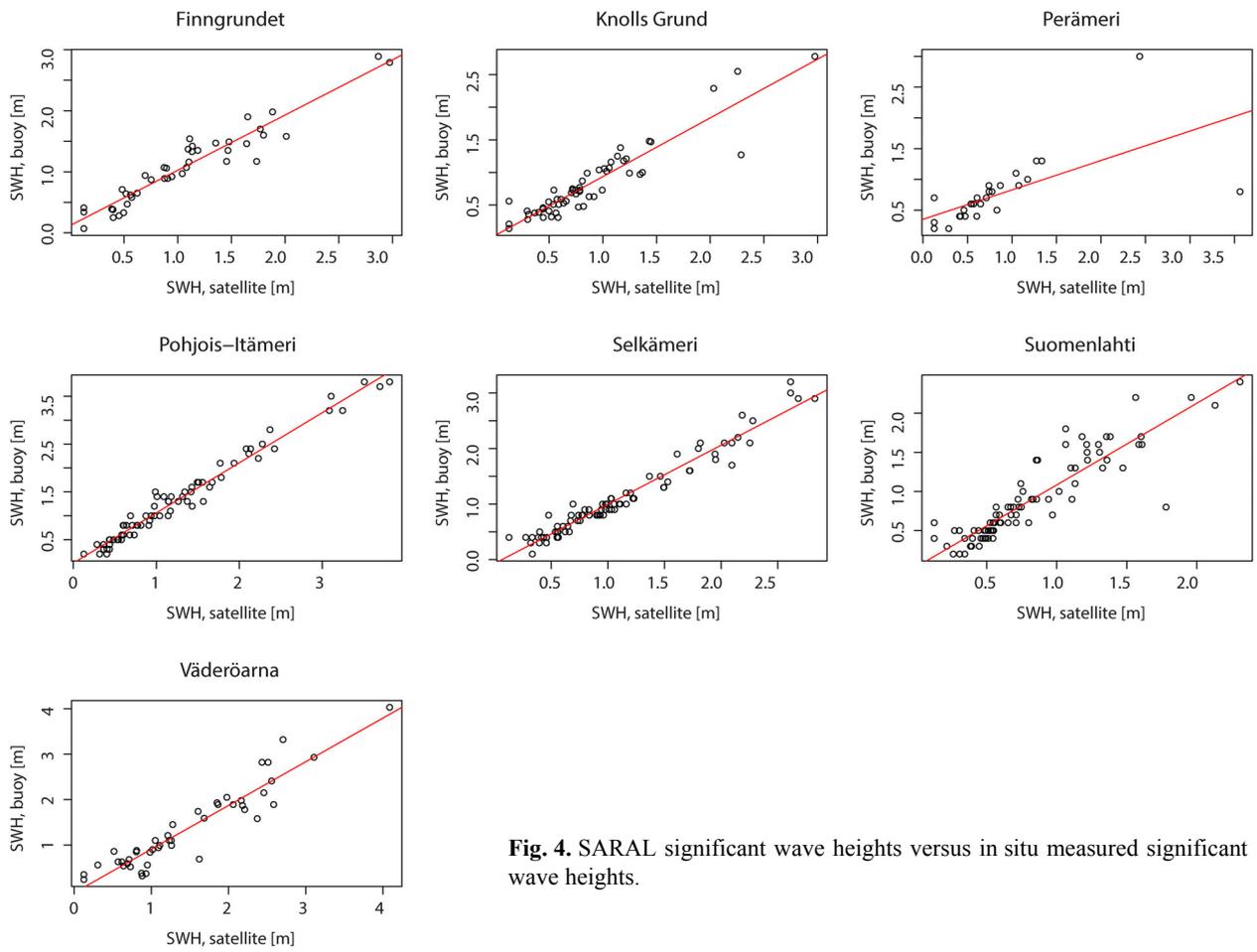

**Fig. 4.** SARAL significant wave heights versus in situ measured significant wave heights.

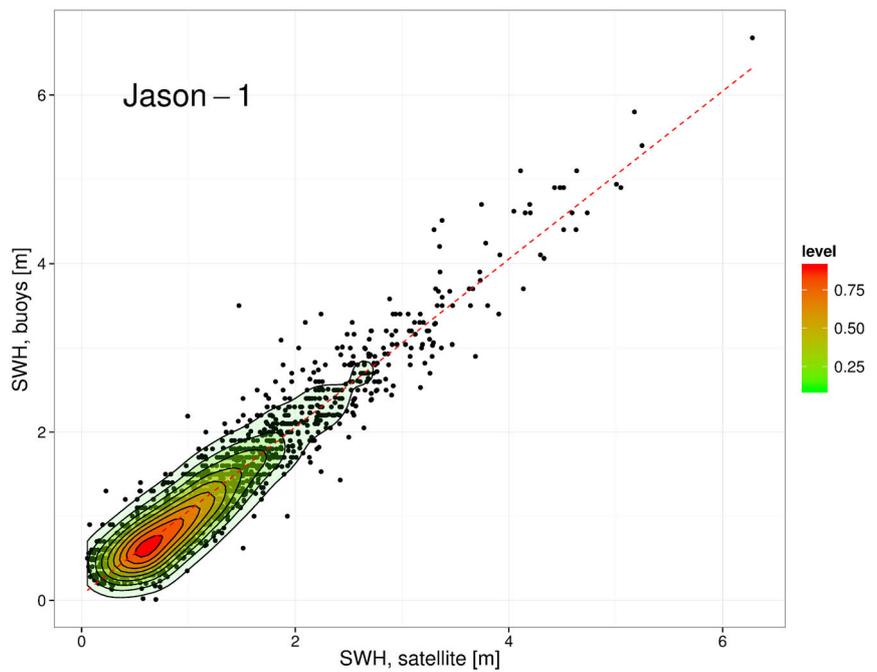

**Fig. 5.** JASON-1 significant wave heights versus all in situ measured wave heights.

Therefore, there is no systematic shift between the wave heights extracted from the JASON-1 data and measured at the sea surface. The slope of the major regression axis is $0.96 \pm 0.01$. The corresponding *p*-value 0.0006 shows that the correlation is significant at a much higher level than 99%. The difference between the ideal match (with a slope equal to 1) and the empirical slope indicates a systematic (albeit fairly small) difference in the two sets of wave heights.

Due to the limited amount of adequate cross-matched data, it was not possible to derive systematic differences from fitting a regression line in the buoy cross-matches for the other satellites. For this reason the relevant match is characterized using the differences between altimeter-measured significant wave heights and in situ measured wave heights using the root-mean-square error (RMSE)

$$\text{RMSE} = \sqrt{\frac{1}{n} \sum (H_{\text{sat}} - H_{\text{buoy}})^2}, \qquad (1)$$

and the average value $\mu$ of differences between the SWH retrieved from satellite altimetry $H_{\text{sat}}$ and from in situ measurements $H_{\text{buoy}}$:

$$\mu = \frac{1}{n} \sum (H_{\text{sat}} - H_{\text{buoy}}). \qquad (2)$$

Here $n$ is the number of cross-matching satellite and buoy data samples. This analysis is complemented by means of fitting the histogram (empirical distribution) of differences of $H_{\text{sat}}$ and $H_{\text{buoy}}$ with a normal distribution (with a mean $\mu$) and finding the standard deviation $\sigma$ of this distribution. A simple estimate of the error (or uncertainty) of the average difference $\mu$ is calculated as $\sigma/\sqrt{n}$ (Table 3).

## VALIDATION OF WAVE INFORMATION FROM SATELLITES

### Deviations of in situ and satellite measurements

The RMSE is less than 0.4 m for all satellites (except for Phase 1 of GEOSAT). This is consistent with the expected uncertainty of 0.3 m for the half an hour time separation between the satellite and buoy observations (Monaldo 1988; Zieger et al. 2009). The Phase 1 of GEOSAT measurements contained initial satellite data in 1985–1989. These data were most likely prone to additional systematic errors that may have been present in the early stage of the satellite era. Unfortunately, the number of cross-matches with the in situ data in the Baltic Sea is insufficient to find out which types of errors are responsible for the large scatter. Owing to the exceptionally large level of RMSE for GEOSAT Phase 1 (>1 m), it is not recommended to use GEOSAT Phase 1 observations for the evaluation of the properties of wave fields and this data set is excluded from further analysis.

Systematic differences between the satellite data and in situ measurements are fairly small for all satellites included in further analysis (Table 3). While TOPEX tends to overestimate the SWH by about 0.17 m on average, ENVISAT, ERS-1 and GEOSAT Phase 2 tend to underestimate the SWH by 0.15–0.23 m. The systematic bias is less than 0.06 m for all other satellites. We also checked for systematic problems in $H_{\text{sat}} - H_{\text{buoy}}$ vs $H_{\text{sat}}$ plots and did not find any dependence on the wave height.

### Distance from the coast

The level of scattering between the altimetry and in situ data depends to some extent on the location of the particular measurement site. The cross-matches between the buoys located closer to the coast (Troubadour, Oskar, Suomenlahti) showed larger scatter between the altimeter data $H_{sat}$ and wave heights $H_{buoy}$ measured by buoys. To understand the reasons for this feature, we calculated the distance from the land for each altimetry snapshot to check how strongly the data were affected by the proximity of the coast. We used the classic digital bathymetry of the Baltic Sea taken from Seifert et al. (2001) with a resolution of about one nautical mile and calculated the distance from the coast for the centroid of each cross-matching area of the altimeter measurement.

**Table 3.** Statistical properties (RMSE, mean $\mu$, standard deviation $\sigma$ of a fitted normal distribution and the number of data points $n$) of cross-matches between satellite altimetry and in situ data. All quantities are presented in metres

| Satellite | RMSE | $\mu$ | $\sigma$ | $n$ |
|---|---|---|---|---|
| CRYOSAT-2 | 0.36 | $0.04 \pm 0.02$ | $0.35 \pm 0.01$ | 582 |
| ENVISAT | 0.33 | $-0.15 \pm 0.03$ | $0.29 \pm 0.02$ | 97 |
| ERS-1 | 0.37 | $-0.18 \pm 0.04$ | $0.32 \pm 0.03$ | 60 |
| ERS-2 | 0.35 | $-0.04 \pm 0.02$ | $0.34 \pm 0.01$ | 378 |
| GEOSAT, phase 1 | 1.17 | $0.3 \pm 0.2$ | $1.1 \pm 0.2$ | 18 |
| GEOSAT, phase 2 | 0.28 | $-0.23 \pm 0.04$ | $0.15 \pm 0.03$ | 82 |
| JASON-1 | 0.26 | $0.056 \pm 0.006$ | $0.256 \pm 0.004$ | 2014 |
| JASON-2 | 0.23 | $0.01 \pm 0.02$ | $0.23 \pm 0.02$ | 117 |
| SARAL | 0.25 | $-0.01 \pm 0.01$ | $0.255 \pm 0.009$ | 412 |
| TOPEX | 0.37 | $0.17 \pm 0.04$ | $0.32 \pm 0.02$ | 157 |

Not unexpectedly, data from all satellites showed a strong dependence of the range measurement uncertainty on the distance from the land. The clearest reliance on the separation from the coast is evident in the GEOSAT data where the data were evidently corrupted when the centroid of the cross-matching area was located less than 0.2° from the coast (Fig. 6). The samples from other satellites were less sensitive with respect to the neighbouring shores; however, a clear increase in the level of scattering in buoy-satellite cross-matches was evident for locations separated by less than 0.2° from the land. It is therefore strongly recommended to exclude the altimeter data closer than 0.2° to any section of the coast in future applications of the altimeter data over the Baltic Sea.

## Ice cover

To validate the ice flags provided in the RADS data set and to study the effect of ice cover on the quality of altimeter data, we calculated ice concentration for each altimeter measurement. The ice concentration measurements (OSI-409-a dataset) were taken from EUMETSAT OSI SAF Global Sea Ice Concentration Reprocessing data (EUMETSAT Ocean and Sea Ice Satellite Application Facility. Global sea ice concentration reprocessing dataset 1978–2015 (v1.2, 2015). Norwegian and Danish Meteorological Institutes, http://osisaf.met.no).

The ice concentration maps were generated with brightness temperature data from the Nimbus-7 Scanning Multichannel Microwave Radiometer and the Defense Meteorological Satellite Program. The data were taken for 37 years during 1978–2015 with a time resolution of one day. For each satellite altimetry measurement, the ice concentration was calculated as an average over a five-kilometre radius, comparable with the footprint of satellite altimetry analysed in the paper (~2–20 km). If no ice data were available for the day, the ice level was flagged as not available.

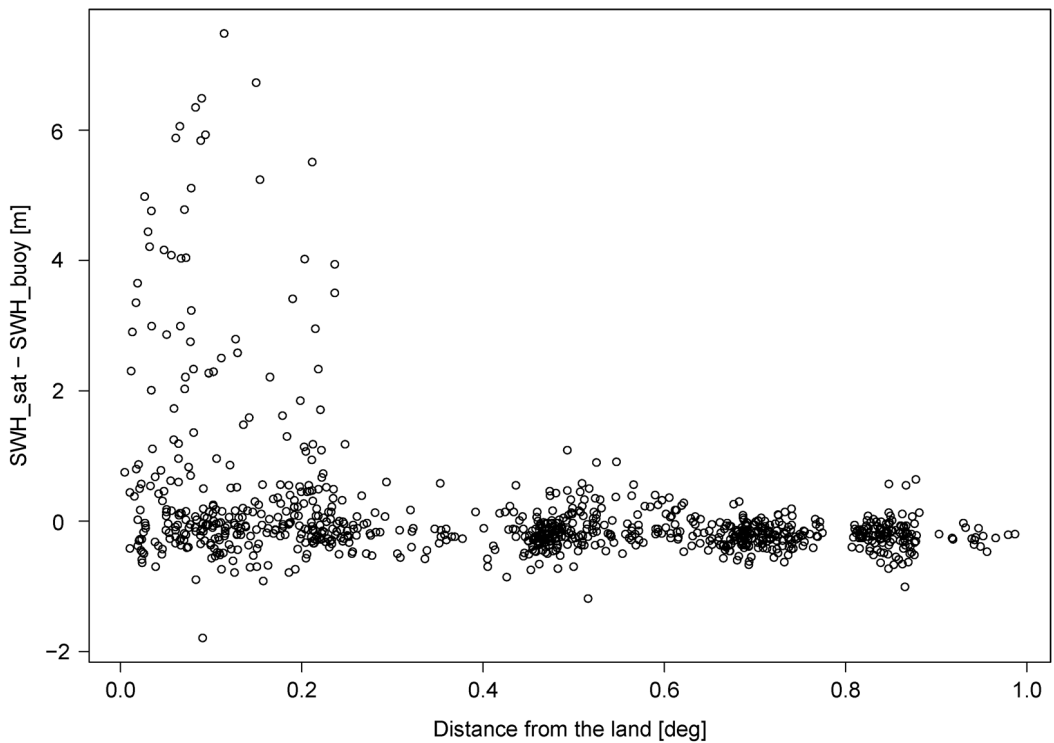

**Fig. 6.** Difference between the GEOSAT significant wave height and in situ data versus ice distance from the land.

The values of SWH retrieved from altimetry snapshots over sea areas with larger ice concentration are characterized by greater uncertainties in the range, backscatter coefficient and wave height. The distribution of SWH exhibits the significant difference between measurements with high and low ice concentration (Fig. 7). The wave heights show, on average, higher values when ice concentration is less than 30%, whereas the ice seems to affect the data starting from as low concentrations as 10%. It is not clear beforehand whether this feature simply reflects the damping effect of waves by floating ice pieces (Wadhams et al. 1988) or whether the results of altimetry are additionally affected by the presence of ice. It is, however, clear that the inclusion of altimeter data measured over the areas with high ice concentrations can affect the overall wave height statistics in the Baltic Sea. It is thus recommended to exclude data from regions with substantial ice coverage from the analysis of wave climate.

A more detailed comparison of the RADS and EUMETSAT data reveals that the ice flag in the RADS data was turned on at the ice concentration of 50%. The flagging itself is highly consistent: no data with an ice concentration of more than 50% missed the ice flag, and this threshold of flagging of ice conditions is appropriate for the Baltic Sea conditions. For example, flagging the data with an ice concentration of >30% results in the additional exclusion of 0.1% of the altimetry data.

**Bias between the satellites**

To look for possible systematic differences between the SWH retrieved using different missions, we performed a cross-matching between satellites. As JASON-1 has the largest amount of cross-matching data with in situ measurements, we focus on the comparison of its data flow with wave heights retrieved from other missions. The procedure was carried out with the same parameters as cross-matching with the buoys, namely, only snapshots that were separated by less than 0.3° and with time difference less than half an hour. The set of data used for cross-matching covers the whole of the Baltic Sea. All the flagged and questionable data were removed from the comparison. Similarly to the above, a regression line was fitted to significant wave height cross-matches to find the systematic deviations. A reduced major axis regression was used to estimate the best fit (Zieger et al. 2009). The regression line modelling results (intercept, slope, *p*-value and significance) are listed in Table 4. As above, we employ the 99% level of statistical significance, that is, set a value as significant if the *p*-value is less than 0.01.

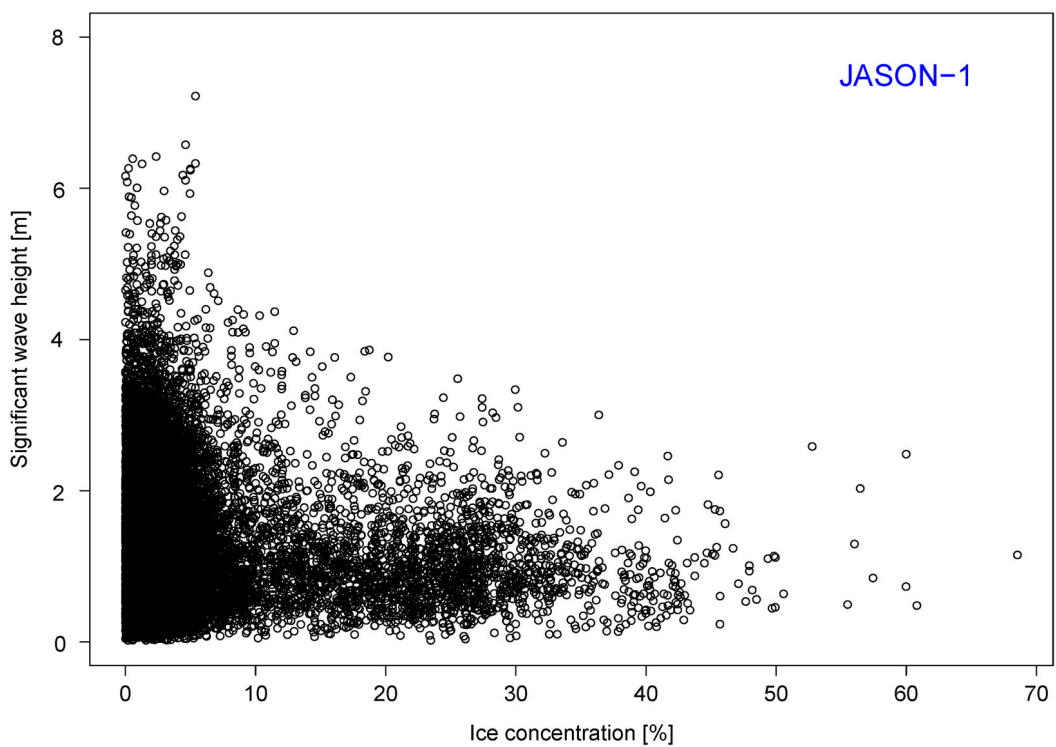

**Fig. 7.** JASON-1 significant wave height versus ice concentration in per cent.

The results are consistent with the comparison of the outcome of different missions with the in situ measurements (see Table 3). For example, the bias between the SWH from CRYOSAT-2 and JASON-1 is $0.04 \pm 0.06$ m, while the bias between the buoys and CRYOSAT-2 data is $0.04 \pm 0.02$ m. The systematic deviation between ENVISAT and JASON-1 is $-0.19 \pm 0.03$ m

(Fig. 8, Table 4), whereas there is a $-0.15 \pm 0.03$ m bias between the ENVISAT satellite and the buoys. The data from all other satellites also show a very consistent pattern of deviations from the JASON-1 data set.

Several reasons such as sensor damage, change of orbit or drift of electronic devices can result in time-variable biases between satellites, which can affect

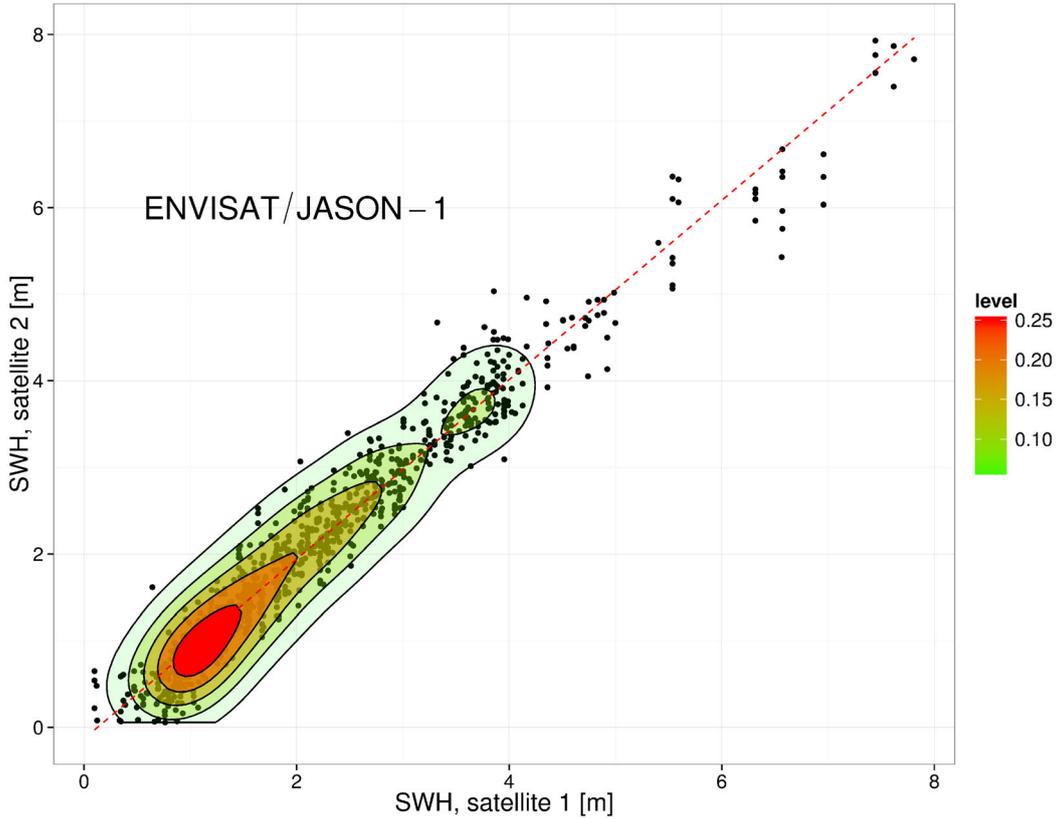

**Fig. 8.** Comparison of significant wave heights from JASON-1 and ENVISAT.

**Table 4.** Properties of fitted regression lines for intercomparison of data from various satellites. Sig. stands for statistical significance at a 99% level and $m_i$ is a modulation index that represents time variability of the bias

| Satellite | Intercept I [m] | I $p$-value | Sig. | Slope S | S $p$-value | Sig. | $m_i$ | $m_i$ [m] |
|---|---|---|---|---|---|---|---|---|
| CRYOSAT-2/JASON-1 | $0.04 \pm 0.06$ | 0.2 | No | $1.04 \pm 0.02$ | 0.03 | No | $49 \pm 4\%$ | 0.02 |
| ENVISAT/JASON-1 | $-0.19 \pm 0.03$ | $1.09 \times 10^{-9}$ | Yes | $1.06 \pm 0.01$ | $1.09 \times 10^{-5}$ | Yes | $71 \pm 7\%$ | 0.13 |
| ERS-1/TOPEX* | $-0.37 \pm 0.04$ | $2.9 \times 10^{-16}$ | Yes | $0.99 \pm 0.02$ | 0.4 | No | $12 \pm 4\%$ | 0.05 |
| ERS-2/JASON-1 | $-0.12 \pm 0.07$ | 0.04 | No | $0.99 \pm 0.04$ | 0.4 | No | $11 \pm 2\%$ | 0.01 |
| GEOSAT/JASON-1 | $-0.15 \pm 0.07$ | 0.02 | No | $0.96 \pm 0.03$ | 0.1 | No | $14 \pm 3\%$ | 0.02 |
| JASON-2/JASON-1 | $0.005 \pm 0.008$ | 0.2 | No | $1.009 \pm 0.004$ | 0.008 | Yes | $62 \pm 1\%$ | 0.01 |
| SARAL/JASON-2* | $-0.02 \pm 0.09$ | 0.4 | No | $1.01 \pm 0.04$ | 0.3 | No | $380 \pm 5\%$ | 0.08 |
| TOPEX/JASON-1 | $0.004 \pm 0.004$ | 0.1 | No | $1.014 \pm 0.002$ | $6.8 \times 10^{-17}$ | Yes | $17 \pm 1\%$ | 0.01 |

* These satellites could not be compared to the JASON-1 data due to lack of simultaneous measurements.

further wave climate studies. To examine whether the satellites considered in this study exhibit this kind of bias changes, we analysed the time variability of biases between satellites. For this purpose, we calculated the mean (bias) and RMSE differences between the SWH retrieved from satellite pairs for each year separately. The level of variation of the bias can be, to a first approximation, quantified using a sort of modulation index $m_i$ (see the last two columns of Table 4):

$$m_i[\%] = 100 \, \mathrm{std}(\mu_i)/<\mu_i>, \qquad (3)$$

where $\mu_i$ is defined in Eq. (2) but is calculated separately for each year $i$, $\mathrm{std} = \sqrt{1/n \sum (x_i - <x>)^2}$ is a standard deviation, $<>$ denotes averaging $<x> = 1/n \sum x_i$ and $n$ is the number of data points $x_i$.

The bias between ERS-2 and JASON-1, GEOSAT and JASON-1, TOPEX and JASON-1, and ERS-1 and TOPEX significant wave height data is of the order of 10%. Consequently, the SWH time series from these satellites contain a very low level of temporal biases. The other satellites showed a significant drift of 50–70% in terms of the modulation index. The highest time changes are between the SARAL and JASON-2 satellites. Even though in this particular case part of the drift can be explained by a different frequency of SARAL observations, the data in Table 4 strongly hint that some of the satellite altimetry data sets are not homogeneous in time. Consequently, as suggested in Queffeulou (2004) and Queffeulou & Croizé-Fillon (2012), in most occasions the time-variable bias between the SWH time series provided by some satellites should be taken into account in further studies using satellite altimetry data for the Baltic Sea. Even though the bias is much lower than in the global ocean (Queffeulou et al. 2010), apparently owing to relatively low overall wave intensity in the Baltic Sea, this suggestion is eventually most relevant for such pairs of satellites as ENVISAT/JASON-1 (0.13 m), SARAL/JASON-2 (0.08 m) and ERS-1/TOPEX (0.05 m), since the absolute changes in bias between these satellites is the largest and reaches ~0.1 m.

## DISCUSSION AND CONCLUSIONS

We presented for the first time a complete cross-validation of multi-mission satellite altimetry data and in situ wave measurements available for the Baltic Sea basin. The focus was on the selection of a reasonable-quality altimetry data subset for the subsequent evaluation of the main properties of the wave climate and its spatio-temporal changes in the Baltic Sea.

The comparison of significant wave heights retrieved from the satellite altimetry data for the Baltic Sea with the available in situ (buoy and echosounder) observations first suggests that the satellite measurements with the backscatter coefficients >13.5 cdb or errors in the SWH normalized standard deviation more than 0.5 m are generally not reliable and should not be used in any analysis of wave properties. The high levels of back-scatter coefficients usually correspond to low waves that generally are unreliably represented by altimetry measurements. Note that the current study could not examine the quality of the altimeter data in the regime of very high waves since the maximum in situ measured SWH in the Baltic Sea is 8.2 m (Tuomi et al. 2011) and SWH in extreme storms hardly exceed 9.5 m (Soomere et al. 2008).

We also examined how the distance from the land and the presence of ice affect the quality of the altimeter data. It was found that altimetry snapshots with centroids closer than 0.2° from the land should be used, if at all, with extreme care. The analysis of the effect of ice concentration on the quality of the altimeter data showed that the ice flag masks the data with the ice concentration of 50% in the common remote sensing data sets. Our comparison indicated that the presence of ice affects the data starting from concentrations around 10%, but substantial effects are evident for concentrations >30%. Even though it is hard to separate the influence of ice concentration on altimeter measurements and the effect of wave attenuation due to ice, we recommend to include in the analysis of wave climate the data with the ice concentration less than 30%.

Cross-matching the altimetry data selected based on these criteria with in situ measurements showed a very good correspondence between these data sets. Only GEOSAT Phase 1 data covering the years 1985–1989 contain substantial deviations from the ground truth and should not be used for wave studies. The root-mean-square difference and bias of altimetry and in situ data are in the range of 0.23–0.37 and ±0.23 m, respectively. The bias for CRYOSAT-2, ERS-2, JASON-1/2 and SARAL is below 0.06 m. A few systematic biases between the altimetry and in situ data sets are consistent with the biases found for the data from different satellites.

As also noted for the global ocean (Queffeulou 2004; Queffeulou & Croizé-Fillon 2012), the altimetry data from several satellites exhibit substantial temporal drift. The magnitude of this drift is much smaller in the Baltic Sea than in the open ocean apparently because of overall lower wave intensity. A large time-variability in the biases was found for the pairs of satellites ENVISAT/JASON-1 (0.13 m), SARAL/JASON-2 (0.08 m) and ERS-1/TOPEX (0.05 m). This feature warns that the relevant time series, most probably, are not homogeneous in time, and their use for wave climate studies might be complicated. However, the biases between the satellites exhibit strong time variability and can change by ~0.1 m, so they should be taken into account in the future studies. The GEOSAT Phase 1 data from 1985–1989 (see Table 1) could not be adequately validated due to the sparseness of early in situ data and lack of the availability of simultaneous satellite observations. This should be taken into account when using the results of the paper.

A new high-resolution (measured at a higher-frequency Ka-band) altimeter data set from the SARAL satellite showed a very good correspondence with the in situ data with no systematic bias (−0.01 ± 0.01 m). Moreover, this data set also showed very good match with the data stream from previous satellites. The bias between the significant wave heights provided by JASON-2 and SARAL is −0.02 ± 0.09 m. The scatter of the relevant data points is fairly limited, and the slope of the relevant regression line is 1.01 ± 0.04 m and its deviation from the ideal value of 1 is not significant. As the SARAL satellite data have a very good coverage of the Baltic Sea (see Fig. 1), the data from this satellite are probably the most valuable asset for the quantification of the Baltic Sea wave climate and understanding of spatio-temporal patterns of its changes.

**Acknowledgements.** The research was funded by the European Economic Area Financial Instrument 2009–2014 Programme 'National Climate Policy', Small Grants Scheme Project 'Effects of Climate Changes on Biodiversity in the Coastal Shelves of the Baltic Sea' 2015–2016 (EEA grant No. 2/EEZLV02/14/GS/022) and supported by the institutional financing by the Estonian Ministry of Education and Research (Grant IUT33-3). We thank A. Giudici for providing the tool for downloading the wave height measurements from the Finnish Meteorological Institute, the anonymous referees for useful comments which improved the article and N. Delpeche-Ellmann and A. Giudici for reviewing the manuscript and helpful comments.

## REFERENCES

Alari, V. 2013. *Multi-Scale Wind Wave Modeling in the Baltic Sea*. PhD Thesis. Marine Systems Institute, Tallinn University of Technology, 134 pp.

Broman, B., Hammarklint, T., Rannat, K., Soomere, T. & Valdmann, A. 2006. Trends and extremes of wave fields in the north-eastern part of the Baltic Proper. *Oceanologia*, **48**(S), 165–184.

Bronner, E., Guillot, A., Picot, N. & Noubel, J. 2013. *SARAL/AltiKa Products Handbook*. Centre National d'Etudes Spatiales, [online]. Available at http://www.aviso.oceanobs.com/fileadmin/documents/data/tools/SARAL_Altika_products_handbook.pdf [viewed 17 July 2016].

Cavaleri, L. & Sclavo, M. 2006. The calibration of wind and wave model data in the Mediterranean Sea. *Coastal Engineering*, **53**(7), 613–627.

Cieślikiewicz, W., Paplińska-Swerpel, B., Kowalewski, M., Bradtke, K. & Jankowski, A. 2008. A 44-year hindcast of wind wave fields over the Baltic Sea. *Coastal Engineering*, **55**(11), 894–905.

Deng, J., Zhang, W., Harff, J., Schneider, R., Dudzinska-Nowak, J., Terefenko, P., Giza, A. & Furmanczyk, K. 2014. A numerical approach for approximating the historical morphology of wave-dominated coasts – A case study of the Pomeranian Bight, southern Baltic Sea. *Geomorphology*, **204**, 425–443.

[ESA] European Space Agency, 2012. *Cryosat Product Handbook*. ESRIN–ESA, Mullard Space Science Laboratory, University College London. Available at https://earth.esa.int/ [viewed 17 July 2016].

Francis, O. P., Panteleev, G. G. & Atkinson, D. E. 2011. Ocean wave conditions in the Chukchi Sea from satellite and in situ observations. *Geophysical Research Letters*, **38**(24), Article number L24610.

Gairola, R. M., Prakash, S., Mahesh, C. & Gohil, B. S. 2014. Model function for wind speed retrieval from SARAL-AltiKa radar altimeter backscatter: case studies with TOPEX and JASON Data. *Marine Geodesy*, **37**( 4), 379–388.

Galanis, G., Hayes, D., Zodiatis, G., Chu, P. C., Kuo, Y.-H. & Kallos, G. 2012. Wave height characteristics in the Mediterranean Sea by means of numerical modeling, satellite data, statistical and geometrical techniques. *Marine Geophysical Research*, **33**(1), 1–15.

Giudici, A. & Soomere, T. 2015. Finnish Meteorological Institute's open data mining tool. In *28th Nordic Seminar on Computational Mechanics, 22–23 October, Tallinn, 2015. Proceedings of the NSCM28* (Berezovski, A., Tamm, K. & Peets, T., eds), pp. 59–62. CENS, Institute of Cybernetics at Tallinn University of Technology, Tallinn.

Harff, J. & Meyer, M. 2011. Coastlines of the Baltic Sea – zones of competition between geological processes and a changing climate: examples from the southern Baltic Sea. In *The Baltic Sea Basin* (Harff, J., Björck, S. & Hoth, P., eds), pp. 149–164. Springer, Heidelberg, Dordrecht, London, New York.

Hithin, N. K., Kumar, V. S. & Shanas, P. R. 2015. Trends of wave height and period in the Central Arabian Sea from 1996 to 2012: a study based on satellite altimeter data. *Ocean Engineering*, **108**, 416–425.

Høyer, J. L. & Nielsen, J. W. 2006. Satellite significant wave height observations in coastal and shelf seas. In *Proceedings of the Symposium on 15 Years of Progress in Radar Altimetry, 13–18 March 2006, Venice, Italy* (Danesy, D., ed.), *ESA Special Publication*, SP-614, Paper No. 812. Noordwijk, The Netherlands.

Hünicke, B., Zorita, E., Soomere, T., Madsen, K. S., Johansson, M. & Suursaar, U. 2015. Recent change – sea level and wind waves. In *Second Assessment of Climate Change for the Baltic Sea Basin* (The BACC II Author Team, ed.), pp. 155–185. Regional Climate Studies, Springer, Cham, Heidelberg.

[JASON] Jason-2, 2008. *Jason-2 Handbook*, SALP-MU-M-OP-15815-CN, Ed 1.2, Aviso Website.

Kahma, K. K. & Calkoen, C. J. 1992. Reconciling discrepancies in the observed growth of wind-generated waves. *Journal of Physical Oceanography*, **22**(12), 1389–1405.

Kudryavtseva, N., Soomere, T. & Giudici, A. 2016. Validation of multi-mission satellite altimetry for the Baltic Sea

region. *Geophysical Research Abstracts*, **18**, EGU2016-5571.

Kumar, U. M., Swain, D., Sasamal, S. K., Reddy, N. N. & Ramanjappa, T. 2015. Validation of SARAL/AltiKa significant wave height and wind speed observations over the North Indian Ocean. *Journal of Atmospheric and Solar-Terrestrial Physics*, **135**, 174–180.

Leppäranta, M. & Myrberg, K. 2009. *Physical Oceanography of the Baltic Sea*. Springer, Berlin, 378 pp.

Liu, Q., Lewis, T., Zhang, Y. & Sheng, W. 2014. Exprimental study on the wave measurements of wave buoys. In *5th International Conference on Ocean Energy, 4–6 November, Halifax*, pp. 1–7. Available at http://www.icoe2014canada.org/wp-content/uploads/2014/11/LiuQuilinARTICLE_18-4.pdf [viewed 17 July 2016].

Martensson, M. & Bergdahl, L. 1987. *On the Wave Climate of the Southern Baltic*. Report Series A:15, Department of Hydraulics, Chalmers University of Technology, Sweden.

Monaldo, F. 1988. Expected differences between buoy and radar altimeter estimates of wind speed and significant wave height and their implications on buoy-altimeter comparisons. *Journal of Geophysical Research*, **93**(C3), 2285–2302.

Nikolkina, I., Soomere, T. & Räämet, A. 2014. Multidecadal ensemble hindcast of wave fields in the Baltic Sea. In *The 6th IEEE/OES Baltic Symposium "Measuring and Modeling of Multi-Scale Interactions in the Marine Environment", May 26–29, Tallinn, Estonia*, pp. 1–9. IEEE Conference Publications.

Orviku, K., Jaagus, J., Kont, A., Ratas, U. & Rivis, R. 2003. Increasing activity of coastal processes associated with climate change in Estonia. *Journal of Coastal Research*, **19**(2), 364–375.

Pettersson, H., Kahma, K. K. & Tuomi, L. 2010. Wave directions in a narrow bay. *Journal of Physical Oceanography*, **40**(1), 155–169.

Pettersson, H., Lindow, H. & Brüning, T. 2013. *Wave Climate in the Baltic Sea 2012*. HELCOM Baltic Sea Environment Fact Sheets 2012. Available at http://helcom.fi/baltic-sea-trends/environment-fact-sheets/hydrography/wave-climate-in-the-baltic-sea/ [viewed 17 July 2016].

Pindsoo, K. & Soomere, T. 2015. Contribution of wave set-up into the total water level in the Tallinn area. *Proceedings of the Estonian Academy of Sciences*, **64**(3), 338–348.

Queffeulou, P. 2004. Long term validation of wave height measurements from altimeters. *Marine Geodesy*, **27**(3–4), 495–510.

Queffeulou, P., Bentamy, A. & Croizé-Fillon, D. 2010. Analysis of seasonal wave height anomalies from satellite data over the global oceans. In *Proceedings of the ESA Living Planet Symposium, 28 June–2 July, Bergen, Norway, SP-686, December 2010*. ESA.

Queffeulou, P. & Croizé-Fillon, D. 2012. *Global Altimeter SWH Data Set*. Technical Report, IFREMER, Brest.

Räämet, A., Suursaar, Ü., Kullas, T. & Soomere, T. 2009. Reconsidering uncertainties of wave conditions in the coastal areas of the northern Baltic Sea. *Journal of Coastal Research, Special Issue*, **56**, 257–261.

R Core Team, 2015. *R: A Language and Environment for Statistical Computing*. R Foundation for Statistical Computing, Vienna, Austria. Available at https://www.R-project.org/ [viewed 17 July 2016].

Rikka, S., Uiboupin, R. & Alari, V. 2014. Estimation of wave field parameters from TerraSAR-X imagery in the Baltic Sea. In *The 6th IEEE/OES Baltic Symposium "Measuring and Modeling of Multi-Scale Interactions in the Marine Environment", May 26–29, Tallinn, Estonia*, pp. 1–6. IEEE Conference Publications.

Ruest, B., Neumeier, U., Dumont, D., Bismuth, E., Senneville, S. & Caveen, J. 2016. Recent wave climate and expected future changes in the seasonally ice-infested waters of the Gulf of St. Lawrence, Canada. *Climate Dynamics*, **46**(1), 449–466.

Ryabchuk, D., Kolesov, A., Chubarenko, B., Spiridonov, M., Kurennoy, D. & Soomere, T. 2011. Coastal erosion processes in the eastern Gulf of Finland and their links with geological and hydrometeorological factors. *Boreal Environment Research*, **16**(Suppl. A), 117–137.

Sandwell, D. T. & McAdoo, D. C. 1988. Marine gravity of the Southern Ocean and Antarctic Margin from Geosat. *Journal of Geophysical Research*, **93**(B9), 10389–10396.

Shaeb, K. H. B, Ananda, A., Joshi, A. K. & Bhandari, S. M. 2015. Comparison of near coastal significant wave height measurements from SARAL/AltiKa with wave rider buoys in the Indian region. *Marine Geodesy*, **38**(1), 422–436.

Scharroo, R. 2012. *RADS version 3.1 User Manual and Format Specifications*. Available at http://rads.tudelft.nl/rads/radsmanual.pdf [viewed 17 July 2016].

Scharroo, R., Leuliette, E. W., Lillibridge, J. L., Byrne, D., Naeije, M. C. & Mitchum, G. T. 2013. RADS: consistent multi-mission products. In *Proceedings of the Symposium on 20 Years of Progress in Radar Altimetry, Venice, 20–28 September 2012. European Space Agency Special Publication*, SP-710, 1–4.

Seifert, T., Tauber, F. & Kayser, B. 2001. A high resolution spherical grid topography of the Baltic Sea. 2nd ed. In *Baltic Sea Science Congress, Stockholm 25–29. November 2001*, Poster 147. Available at www.io-warnemuende.de/iowtopo [viewed 17 July 2016].

Soomere, T. 2016. Extremes and decadal variations in the Baltic Sea wave conditions. In *Extreme Ocean Waves* (Pelinovsky, E. & Kharif, C., eds), pp. 107–140. Springer.

Soomere, T. & Räämet, A. 2011. Spatial patterns of the wave climate in the Baltic Proper and the Gulf of Finland. *Oceanologia*, **53**(1-TI), 335–371.

Soomere, T. & Räämet, A. 2014. Decadal changes in the Baltic Sea wave heights. *Journal of Marine Systems*, **129**, 86–95.

Soomere, T., Behrens, A., Tuomi, L. & Woge Nielsen, J. 2008. Wave conditions in the Baltic Proper and in the Gulf of Finland during windstorm Gudrun. *Natural Hazards and Earth System Science*, **8**(1), 37–46.

Soomere, T., Weisse, R. & Behrens, A. 2012. Wave climate in the Arkona basin, Baltic Sea. *Ocean Science*, **8**, 287–300.

Suursaar, Ü. 2013. Locally calibrated wave hindcasts in the Estonian coastal sea in 1966–2011. *Estonian Journal of Earth Sciences*, **62**(1), 42–56.

Suursaar, Ü. 2015. Analysis of wave time series in the Estonian coastal sea in 2003–2014. *Estonian Journal of Earth Sciences*, **64**(4), 289–304.

Suursaar, Ü., Alari, V. & Tõnisson, H. 2014. Multi-scale analysis of wave conditions and coastal changes in the north-eastern Baltic Sea. *Journal of Coastal Research*, **70**, 223–228.

Taylor, P. K., Dunlap, E., Dobson, F. W., Anderson, R. J. & Swail, V. R. 2002. On the accuracy of wind and wave measurements from buoys. In *Presentations at the DBCP Technical Workshop "Developments in Buoy Technology, Communications, Science and Data Applications"*, Perth, Australia, October 22–23, 2001. DBCP Technical Document No. 21, 15 pp.

Tõnisson, H., Suursaar, Ü., Orviku, K., Jaagus, J., Kont, A., Willis, D. A. & Rivis, R. 2011. Changes in coastal processes in relation to changes in large-scale atmospheric circulation, wave parameters and sea levels in Estonia. *Journal of Coastal Research*, Special Issue **64**, 701–705.

Tuomi, L., Kahma, K. K. & Pettersson, H. 2011. Wave hind-cast statistics in the seasonally ice-covered Baltic Sea. *Boreal Environment Research*, **16**( 6), 451–472.

Tuomi, L., Kahma, K. K. & Fortelius, C. 2012. Modelling fetch-limited wave growth from an irregular shoreline. *Journal of Marine Systems*, **105**, 96–105.

Tuomi, L., Pettersson, H., Fortelius, C., Tikka, K., Björkqvist, J.-V. & Kahma, K. K. 2014. Wave modelling in archipelagos. *Coastal Engineering*, **83**, 205–220.

Vignudelli, S., Kostianoy, A. G., Cipollini, P. & Benveniste, J. (eds). 2011. *Coastal Altimetry*. Springer, Heidelberg, 566 pp.

Viška, M. & Soomere, T. 2012. Hindcast of sediment flow along the Curonian Spit under different wave climates. In *Proceedings of the IEEE/OES Baltic 2012 International Symposium "Ocean: Past, Present and Future. Climate Change Research, Ocean Observation & Advanced Technologies for Regional Sustainability"*, May 8–11, Klaipėda, Lithuania, pp. 1–7. IEEE Conference Publications.

Wadhams, P., Squire, V. A., Goodman, D. J., Cowan, A. M. & Moore, S. C. 1988. The attenuation rates of ocean waves in the marginal ice zone. *Journal of Geophysical Research*, **93**(C6), 6799–6818.

Young, I. R., Zieger, S. & Babanin, A. V. 2011. Global trends in wind speed and wave height. *Science*, **332**(6028), 451–455.

Zhang, W., Deng, J., Harff, J., Schneider, R. & Dudzinska-Nowak, J. 2013. A coupled modeling scheme for long-shore sediment transport of wave-dominated coasts – A case study from the southern Baltic Sea. *Coastal Engineering*, **72**, 39–55.

Zieger, S., Vinoth, J. & Young, I. R. 2009. Joint calibration of multiplatform altimeter measurements of wind speed and wave height over the past 20 years. *Journal of Atmospheric and Oceanic Technology*, **26**(12), 2549–2564.

## Satelliitaltimeetriavahenditega ja kontaktmeetodil mõõdetud Läänemere lainekõrguste võrdlus

### Nadezhda A. Kudryavtseva ja Tarmo Soomere

On esitatud satelliitaltimeetriavahenditega mõõdetud Läänemere oluliste lainekõrguste võrdlus kontaktmõõtmiste tulemustega. On analüüsitud kõigi kümne selleks seni kasutatud tehiskaaslase andmestikke eesmärgiga eristada laine-kliima analüüsiks sobivad mõõtmised. Mitteusaldusväärseteks peetakse mõõtmisi, mille puhul on tagasihajumis-koefitsient suurem kui 13,5, oluline lainekõrguse normaliseeritud standardhälve ületab 0,5 m või mõõteala keskpunkt paikneb rannale lähemal kui 0,2°. Jää mõjutab mõõtmistulemusi veidi juba siis, kui merel on jääd 10%, kuid oluline mõju altimeetria lugemile ilmneb alates 30-protsendilisest jää kontsentratsioonist.

Kirjeldatud moel eristatud lainekõrguse üksikmõõtmised on suhteliselt heas kooskõlas kontaktmõõtmiste tule-mustega üheksa tehiskaaslase puhul. Vaid GEOSAT 1 (1985–1989) andmestiku võrdlemiseks ei ole piisavalt kontakt-mõõtmisi. Altimeetria lugemite ja kontaktmõõtmistest määratud lainekõrguste ruutkeskmise erinevus on 0,23–0,37 m, mis on Läänemere keskmise lainekõrguse (~1 m) taustal siiski arvestatav erinevus. Erineval moel leitud laine-kõrguste süstemaatiline nihe on alla 0,06 m CRYOSAT-2, ERS-2, JASON-1/2 ja SARAL-i puhul, kuid suurem (kuni 0,23 m) ENVISAT-i, ERS-1, GEOSAT-i ja TOPEX-i puhul. Erinevatelt tehiskaaslastelt mõõdetud andmestike (ENVISAT/JASON-1, SARAL/JASON-2, ERS-1/TOPEX) võrdlemisel ilmneb mõnedes andmestikes märkimisväärne ajaline triiv, mis vajab lainekliima uuringute jaoks korrigeerimist. Uusimate tehiskaaslaste (näiteks SARAL) andmes-tikud on parema lahutusvõimega ja kontaktmõõtmiste tulemustega väga heas kooskõlas.